\documentclass[aps,prb,preprint,groupedaddress,showpacs,amsmath]{revtex4}

\usepackage{amsfonts}
\usepackage{graphicx}
\begin{document}

\title{Co/Nb/Co low field superconducting spin switch}

\author{G. Carapella}
\thanks{Corresponding author}
\email[\newline e-mail: ]{giocar@sa.infn.it}
 \affiliation{CNR-INFM
"SUPERMAT"   and Dipartimento di Matematica e Informatica,
Universit\'{a} di Salerno, via Ponte don Melillo, I-84084
Fisciano,Italy.}
\author{F. Russo}
\affiliation{CNR-INFM "SUPERMAT"  and Dipartimento di Matematica e
Informatica, Universit\'{a} di Salerno, via Ponte don Melillo,
I-84084 Fisciano,Italy.}

\author{G. Costabile}
\affiliation{CNR-INFM "SUPERMAT"  and Dipartimento di Matematica e
Informatica, Universit\'{a} di Salerno, via Ponte don Melillo,
I-84084 Fisciano,Italy.}

\date{\today}

\begin{abstract}

We report experiments on a superconducting spin switch based on
technologically relevant materials as elemental ferromagnetic Co and
elemental superconducting Nb. The Co/Nb/Co structure exhibits
inverse spin switch effect, can be operated at liquid helium
temperature and can switch from superconductive to normal state in
rather weak applied magnetic fields. Relevant critical currents as a
function of temperature and magnetic field as well as preparation of
superconductive or resistive state are addressed here.

\end{abstract}


\maketitle

In a Ferromagnet/Superconductor/Ferromagnet (FSF) spin switch,
superconductivity can be controlled by the relative orientation of
the magnetizations of the outer ferromagnetic electrodes sandwiching
the superconductor.  Recently, spin switches based on proximity
coupled metallic ferromagnets \cite{buzdin,tagirov} and classic
metal superconductors \cite{gu,potenza,moraru1,rusanov1,singh} or
high $\rm T_{c}$ superconductors  \cite{pena} have been widely
investigated. The experimental results  suggest that
superconductivity can be depressed both in the parallel ($\rm |P>$)
state of the magnetizations (standard spin switch effect) and in the
antiparallel ($\rm |AP>$) state (inverse spin switch effect). As a
general trend, it seems that standard spin switch effect is observed
when exchange biasing is used \cite{gu,potenza,moraru1} to achieve
the antiparallel state, while the inverse effect is observed when
antiparallel orientation is achieved using different coercive fields
\cite{rusanov1,singh,pena} for the ferromagnetic materials.

Most of the experiments reported in the literature focused on the
individuation of two critical temperatures corresponding to the $\rm
|AP>$ or $\rm |P>$ state of the spin switch, an important issue from
the  point of view of fundamental physics. Here, beside the use of
two technologically relevant materials, as Nb and Co, we report a
complementary experimental study, that may be of some interest for
 applied physics. In particular, we report
current-voltage curves and critical currents as a function of
magnetic field and temperature, as well as a test on preparation of
superconductive or resistive state  at liquid helium temperature.

\begin{figure}[!hbt]
\begin{center}
\includegraphics[width=6.5 cm]{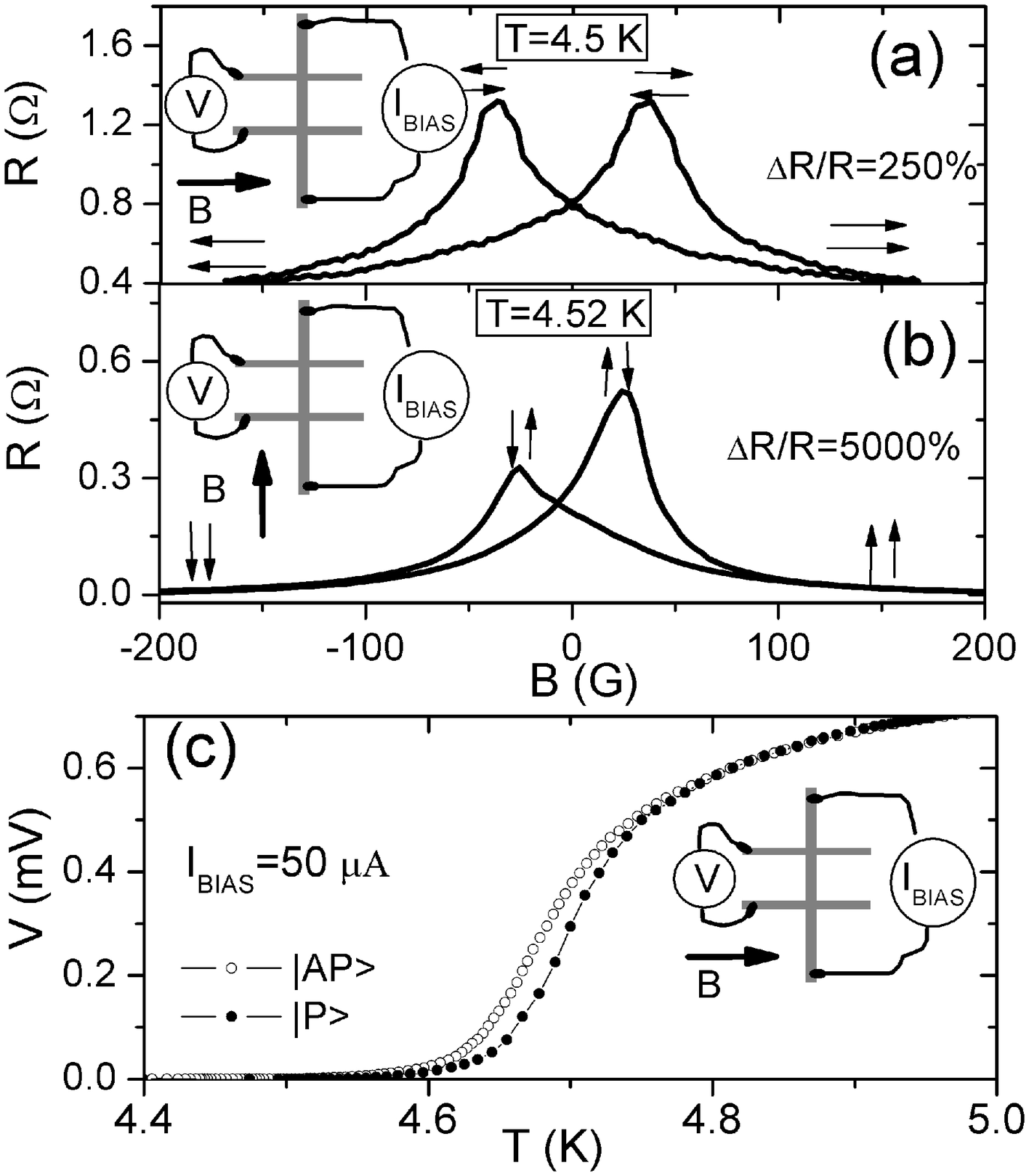}
\caption{Magnetoresistance of the trilayer at a temperature slightly
larger than $\rm T_{c}$ for in plane magnetic field applied
perpendicularly [(a)] or longitudinally [(b)] to the bias current
direction. The arrows show the direction of the magnetizations. (c)
Voltage at fixed bias current versus temperature at two fixed
magnetic fields promoting $\rm |P>$ or $\rm |AP>$
states.}\label{Fig:1}
\end{center}
\end{figure}

The Co/Nb/Co trilayers   were deposited onto glass substrates by rf
magnetron sputtering in a high vacuum system with a base pressure of
$ 2 \times 10^{-7} $ Torr in pure Argon pressure at $3.1 \times
10^{-3}$ Torr at room temperature. Co was deposited at the rate of
0.1 nm/s, producing a roughness smaller than 0.5 nm. The  Nb was
deposited with the rate of  2.2 nm/s, producing a  roughness below
1.5 nm, while affected by very low contamination from magnetic
materials present in the chamber. To achieve different coercive
fields, the bottom and the top Co films were deposited with
different thicknesses, in a pseudo spin valve configuration
\cite{rusanov1,singh,pena}. In all the devices the bottom Co layer
is 8 nm thick, the top Co layer is 16 nm thick, while Nb thickness
ranges from 18 nm to 32 nm. The trilayer is covered by a 2 nm Al cap
layer to prevent the oxidation of the top layer. The whole trilayer
was patterned by photolithography and lift-off in a Hall geometry,
as sketched in the insets of Fig. \ref{Fig:1}, that allows four
contacts measurements. The width of the strips is $\rm 200\ \mu m$
and voltage contact are 3 mm apart.
 All samples showed similar behavior, with a zero resistance critical
 temperature decreasing as  Nb spacer thickness was decreased.
 Here we report about a
sample with 30 nm thick Nb, that allows us to operate easily in a
standard liquid helium cryostat.
\begin{figure}[!hbt]
\begin{center}
\includegraphics[width=6.5 cm]{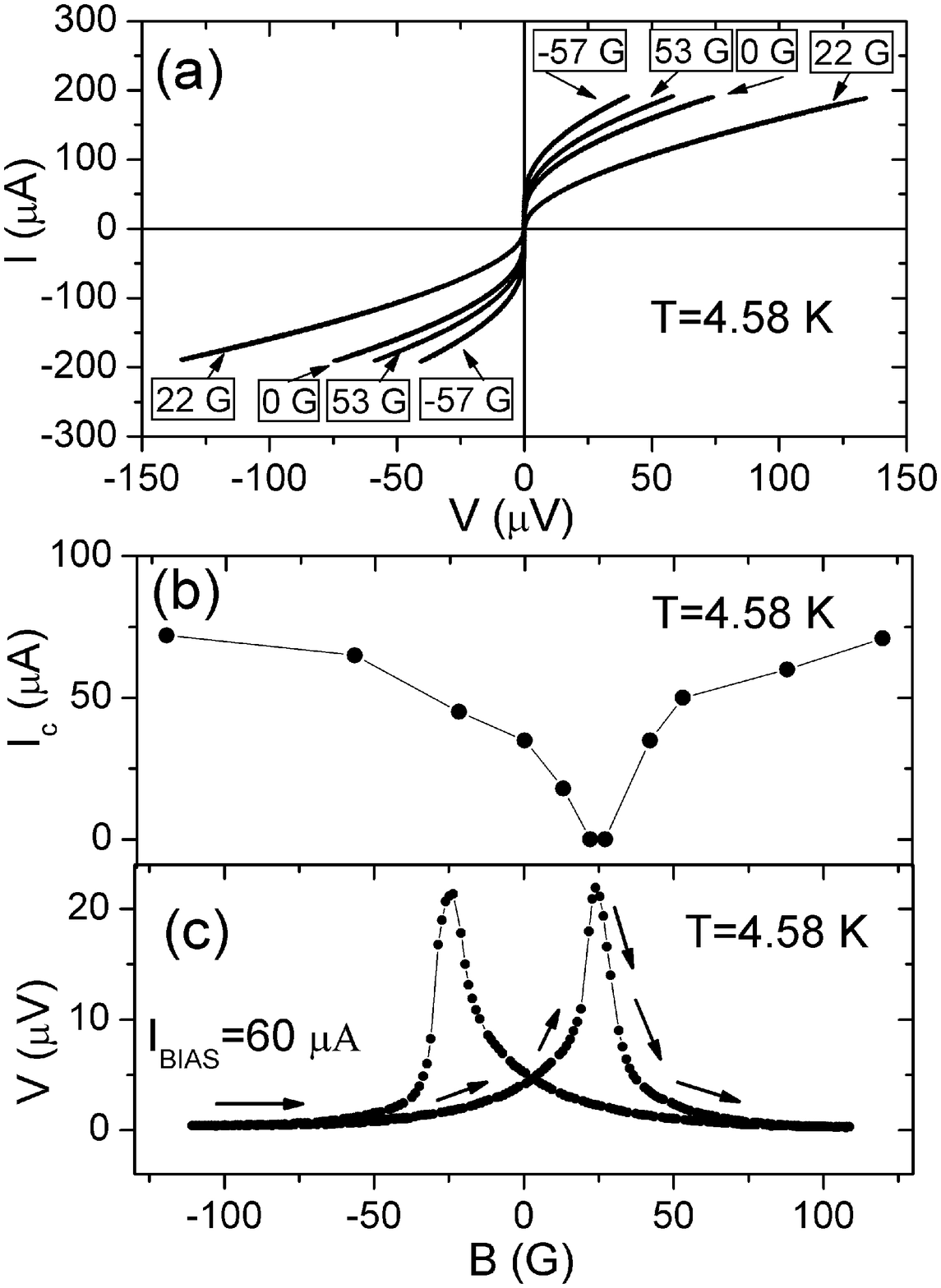}
\caption{(a) Current-Voltage characteristics of the spin switch at
different applied magnetic fields.
 (b) Critical current as a function of magnetic field
  for fields increasing from negative
to positive. (c) Voltage at a fixed bias current vs magnetic field.
The arrows identify the curve traced in the the same magnetic field
half-loop as in (b).} \label{Fig:2}
\end{center}
\end{figure}

In the top panels of Fig. \ref{Fig:1} we show the magnetoresistance
curves of the spin switch at a temperature slightly larger than the
zero resistance critical temperature. The magnetic field is applied
in the plane of the trilayer and (see insets), it is perpendicular
to the current flow in Fig. \ref{Fig:1}(a) and longitudinal to the
current flow in Fig. \ref{Fig:1}(b). The magnetoresistance ratio
$\rm \Delta R/R$ is found very large, of the order of the one
observed \cite{pena} in spin switch based on a highly polarized
ferromagnet. At temperatures where the Nb is normal (above 6 K) the
magnetoresistance curves (not shown) exhibit peaks of resistance
approximately at same locations as in the Fig. \ref{Fig:1},
 but  a vanishingly small $\rm \Delta R/R \approx 0.09\%$,
 as we should expect for a pseudo spin valve
 with a rather thick
 normal spacer layer. Coherently with a pseudo spin valve behavior
 in the normal state,
 we can expect that relative orientations
 for magnetizations of outer electrodes stay the same at
 lower temperature and are directed as
 shown by the arrows in the Fig. \ref{Fig:1}.
Resistance is found to be larger in the $\rm |AP>$ state, i.e., an
inverse spin
 switch effect is exhibited by the Co/Nb/Co trilayer, in agreement
 with other reported spin switch where the pseudo spin valve
 configuration
\cite{rusanov1,singh,pena}
 was used to achieve the antiparallel state.
 Moreover, the absence of any dips
\cite{rusanov2}
 in the curve of Fig. \ref{Fig:1}(b)
 and the presence of well defined peaks in the resistance for both the
 orientations of
 the in plane magnetic field strongly suggest that the whole trilayer
 is operated, as opposite to a  FS bilayer  effect \cite{rusanov2} .
 In Fig. \ref{Fig:1}(c) we show curves of the
 voltage at constant current (proportional to resistance)
 versus temperature
 for two values of the magnetic field that induce an
$\rm |AP>$ or a $\rm |P>$ state for magnetizations. The field is
applied perpendicularly to the current direction, as shown in the
inset, but similar results were obtained  applying the field
longitudinally to the current flow. V-T curves apparently differ of
about 30 mK in the middle of transition. This difference is of the
same order of magnitude of the ones reported
\cite{moraru1,rusanov1,singh} for other spin switches. Moreover, as
expected from proximity effect, the zero voltage critical
temperature of the trilayer ($\rm T_{c}^{FSF}\approx 4.5\ K$ when
$\rm I_{BIAS}=50 \mu\ A$) is found lower than the critical
temperature of a single Nb film having the same thickness, that we
measured as $\rm T_{c}^{S}\approx 6\ K$. The behavior of the spin
switch was found qualitatively the same for both the directions of
the magnetic field with respect to the current direction, therefore
in the following discussion we shall consider the magnetic field
applied perpendicularly to the current.

In Fig. \ref{Fig:2}(a) we show the $I-V$ curves of the spin switch
at different magnetic fields,  recorded at 4.58 K. The curves are
modulated by magnetic fields much weaker than the parallel critical
field of the  Nb film, that at this temperature is estimated to be
several thousands G. The critical current as a function of magnetic
field (increased from negative to positive fields) is shown in Fig.
\ref{Fig:2}(b). The critical current in the $\rm |P>$ state $\rm
I_{c}^{|P>}$, achieved for high negative or positive fields [$\rm
B=\pm  100\ G$ in Fig. \ref{Fig:2}(b)], is much larger than the
critical current in $\rm |AP>$ state $\rm I_{c}^{|AP>}$ , achieved
at about 30 G. Biasing the device with current in-between the two
relevant critical currents, and sweeping the field up and down, the
Voltage $\emph vs$ Field curve shown in Fig. \ref{Fig:2}(c) is
recorded, evidencing a clear transition from the zero-voltage state
to the resistive state. The arrows in Fig. \ref{Fig:2} (c) identify
the branch traced in the same magnetic field half-loop used to
record data in panel (b).
\begin{figure}[!hbt]
\begin{center}
\includegraphics[width=6.5 cm]{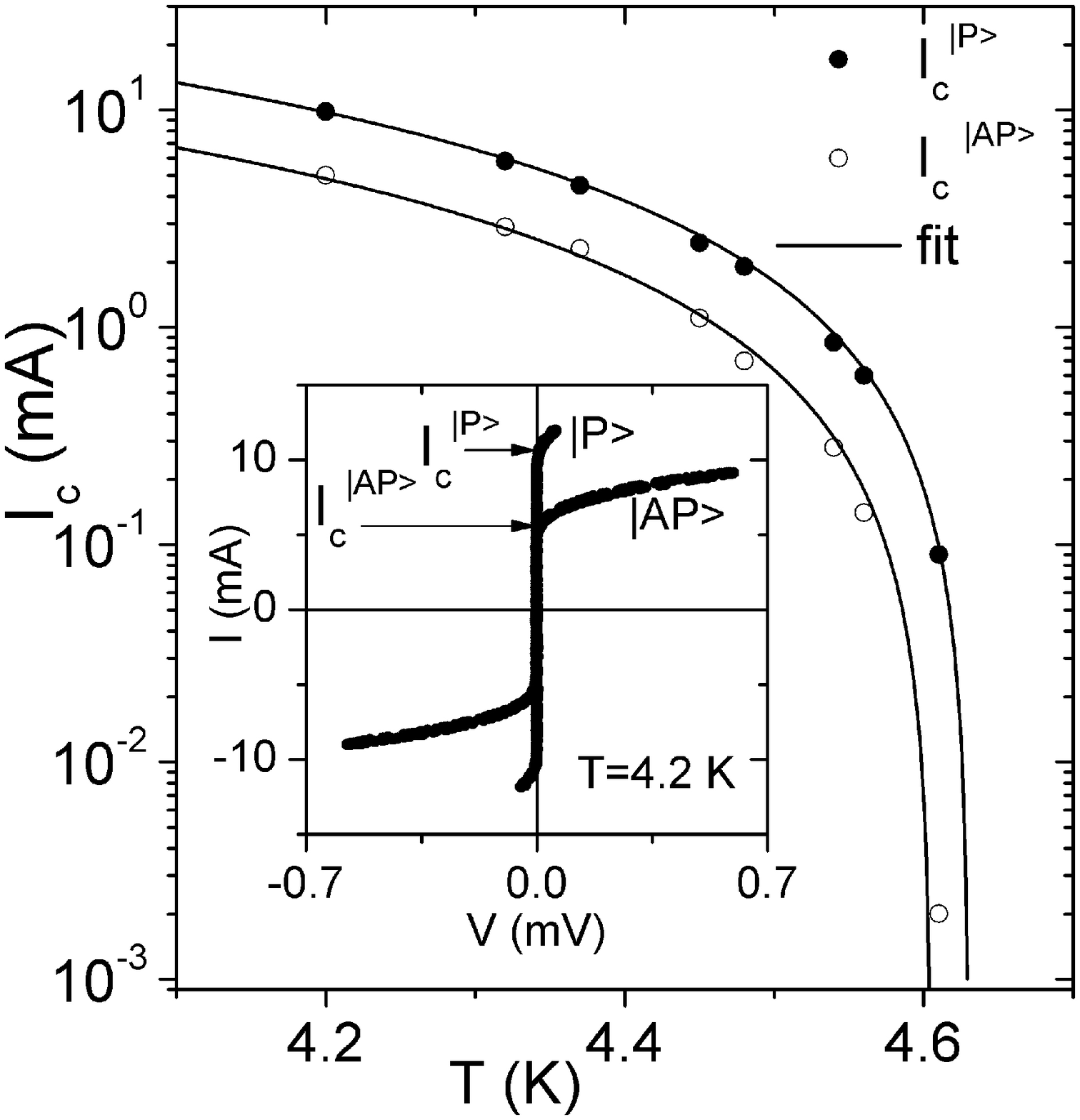}
\caption{ Critical current versus temperature in the $\rm |P>$  or
$\rm |AP>$  state. The solid line is a fit with theory. In the inset
it is shown the $I(V)$ of the spin switch prepared in the two states
recorded at 4.2 K . }\label{Fig:3}
\end{center}
\end{figure}
\begin{figure}[!htb]
\begin{center}
\includegraphics[width=6.5 cm]{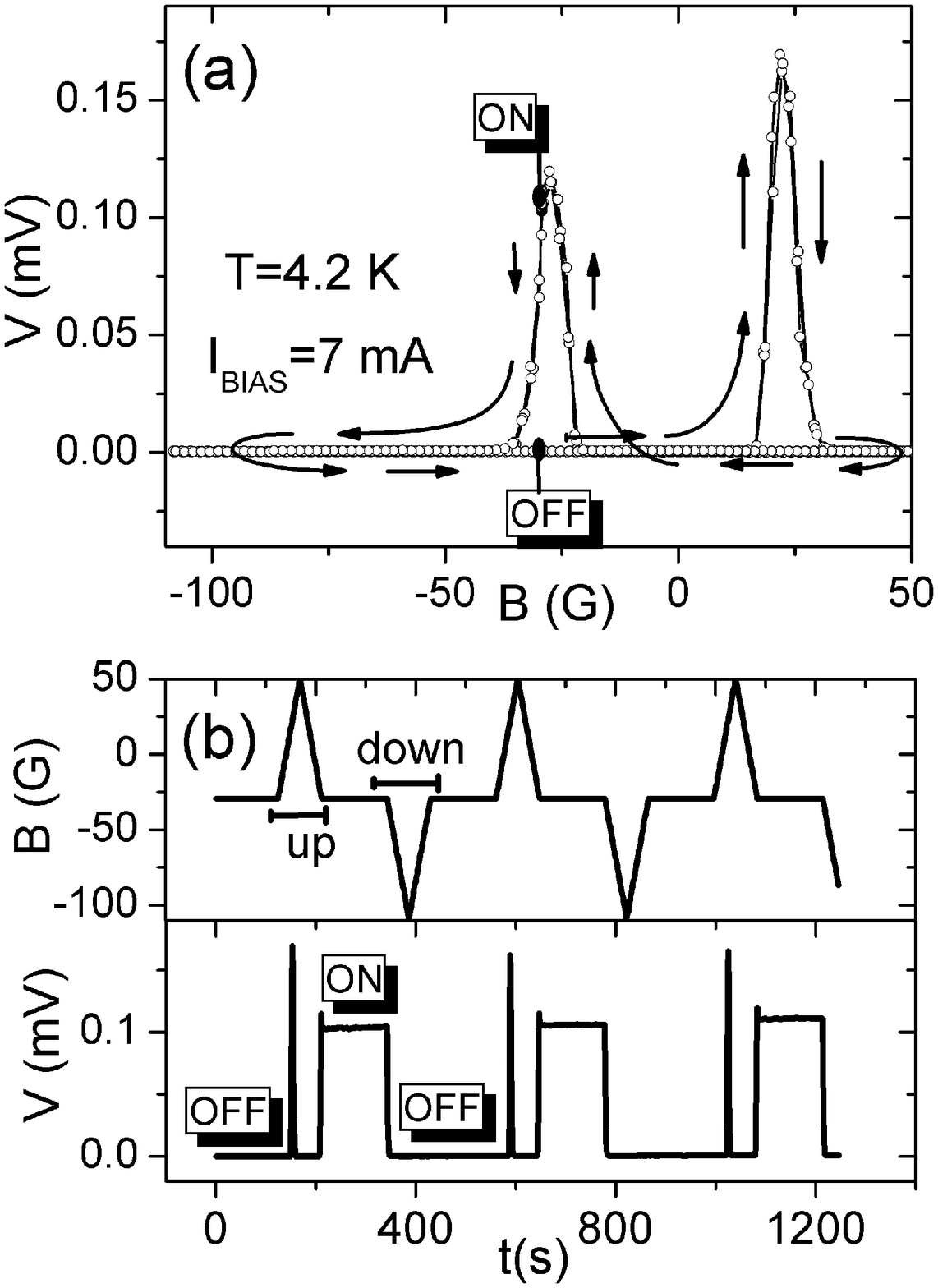}
\caption{ (a) Voltage versus Field curve during a loop of the
magnetic field at 4.2 K. The biasing current is chosen in between
the critical currents of the $\rm |AP>$ and $\rm |P>$ states at the
working temperature. (b) Magnetic field waveform used to prepare the
ON or OFF states together with the Voltage waveform of the spin
switch.}\label{Fig:4}
\end{center}
\end{figure}

The Current-Voltage curves of the spin switch prepared in the $\rm
|P>$ or $\rm |AP>$ state recorded at liquid helium temperature are
shown in the inset of Fig. \ref{Fig:3}, together with the
identification of the two critical currents $\rm I_{c}^{|P>}$ and
$\rm I_{c}^{|AP>}$ at this temperature. From data we estimate a
critical current density in the parallel state at 4.2 K $\rm
J_{c}^{|P>}\simeq 2\times 10^{5}\ A/cm^{2}$.
 The two critical currents as a function of temperature
are reported in the Fig. \ref{Fig:3}. The two currents are always
appreciably different, with $\rm I_{c}^{|P>}$ about doubling $\rm
I_{c}^{|AP>}$ at 4.2 K and also much larger than $\rm I_{c}^{|AP>}$
near the transition temperatures.   Both critical currents are
adequately fitted with the formula \cite{Tink} valid for isolated
thin films
 near the transition temperature
 \begin{displaymath}
I_{c}^{i}=I_{c0}^{i}[1-T/T_{c}^{i}]^{3/2}
\end{displaymath}
where $\rm i=|P>, |AP>$. From the fit we estimated $\rm
T_{c}^{|P>}=4.63 \ K$ and $\rm T_{c}^{|AP>}=4.60\ K$. Moreover, the
critical current densities at $\rm T=0$ can be extrapolated as $\rm
J_{c0}^{|P>}= 5.7\times 10^{6}\ A/cm^{2}$ and $\rm
J_{c0}^{|AP>}=3.1\times 10^{6}\ A/cm^{2}$. So, although a weakening
due to the proximity effect is unavoidable,
 the spin switch is found capable to operate with  critical currents
of almost the same order of magnitude as the critical current of the
isolated superconductive film, as it was theoretically predicted
\cite{tagirov}.
 The fact that in the magneto-quenched state, i.e., our $\rm |AP>$
 state, the
critical current  follows again  a genuine thin superconducting film
behavior allows us to rule out mechanisms of local weakening of
superconductivity (e.g., the ones caused by fringe field effects
discussed by Clinton and  Johnson \cite{johnson}). Moreover, in the
framework in which the above formula is derived \cite{Tink}, the
critical current is proportional to the squared superconducting
energy gap. So data in Fig. \ref{Fig:3} suggest that an appreciable
gap suppression is associated to the $\rm |AP>$ state of our spin
switch.

As can be inferred from Fig. \ref{Fig:3} and Fig. \ref{Fig:2},
biasing the device with a constant current $\rm
I_{c}^{|AP>}<I_{BIAS}<I_{c}^{|P>}$, a transition from zero-voltage
state to the resistive state and viceversa can be achieved at all
temperatures below the critical ones during a cycle of the applied
magnetic field. This is shown in Fig. \ref{Fig:4}(a) for our device
operated at 4.2 K while biased with 7 mA. The arrows indicate a
single cycle of the applied magnetic field. Depending on history, at
the same magnetic field can correspond two different voltage levels.
Choosing properly the bias current, these voltage levels can either
be $\rm V=0$ or $\rm V\neq0$, labeled as ON and OFF states in the
Fig. \ref{Fig:4}(a). In an application as a bit, it is convenient to
choose the voltage of the ON state as large as possible, i.e., the
magnetic field should take the value that produce $\rm |AP>$ state.
An example of magnetic field waveform that can be used to set the
two states is shown in Fig. \ref{Fig:4}(b), where we show also the
time trace of the Voltage measured across the spin switch. The
upward section of the $\rm B(t)$, labeled $\emph up$ in
\ref{Fig:4}(b), prepares the ON state, while the downward section,
labeled $\emph down$, prepares the OFF state. After preparation, the
device stay stable in the states. In the \ref{Fig:4}(b) the two
states are prepared in a sequence, as an example.  We used the
magnetic field generated by a superconductive solenoid to build the
$\rm B(t)$ waveform, but, due the very low magnetic fields involved,
an insulated superconductive control line deposited on the top of
the spin switch could do as well.

 The inverse spin switch effect reported here  is possibly
  accounted for
gap suppression \cite{rusanov1,pena,singh} by means of spin
imbalance \cite{takahashi} in the $\rm |AP>$ state, since  the spin
polarization of the metallic Co ($\rm P\approx 0.4$) is relatively
large and Nb thickness here is close to its spin diffusion length
(estimated \cite{singh} around 30 nm for Nb at cryogenic
temperature).

 Summarizing, we reported inverse spin switch
effect in a Co/Nb/Co trilayer at liquid helium temperature. To the
parallel and antiparallel state correspond two appreciably different
critical currents pointing at two slightly different critical
temperatures. This suggests a superconductive energy gap suppression
in the antiparallel configuration of magnetizations.
 Biasing the device with a constant current comprised between the two relevant
critical currents, the preparation of the device in the
superconductive or resistive state can be reliably achieved using
weak magnetic fields.

\end{document}